\documentclass[3p]{elsarticle}
\journal{Signal Processing}

\usepackage{amsmath,amssymb,bm}
\usepackage{subfigure}
\usepackage{psfrag,graphicx,color}
\usepackage{url}
\usepackage[normalem]{ulem}

\graphicspath{{imgs/}}

\definecolor{dgreen}{rgb}{0,.6,0}

\begin{document}

\begin{frontmatter}

\title{Cryptanalysis of a family of self-synchronizing chaotic stream ciphers}

\author[Spain]{David Arroyo\corref{corr}}
\author[Spain]{Gonzalo Alvarez}
\author[Spain2]{Jos\'{e} Mar\'{\i}a Amig\'{o}}
\author[germany]{Shujun Li}
\cortext[corr]{Corresponding author: David Arroyo
(david.arroyo@iec.csic.es).}
\address[Spain]{Instituto de F\'{\i}sica Aplicada, Consejo Superior de
Investigaciones Cient\'{\i}ficas, Serrano 144, 28006 Madrid, Spain}
\address[Spain2]{Centro de Investigaci\'on
Operativa, Universidad Miguel Hern\'andez, Avda. de la Universidad
s/n, 03202 Elche, Spain}
\address[germany]{Fachbereich Informatik und Informationswissenschaft,
Universit\"{a}t Konstanz, \\Fach M697, Universit\"{a}tsstra{\ss}e
10, 78457 Konstanz, Germany}

\begin{abstract}
Unimodal maps have been broadly used as a base of new encryption
strategies. Recently, a stream cipher has been proposed in the
literature, whose keystream is basically a symbolic sequence of the
(one-parameter) logistic map or of the tent map. In the present work
a thorough analysis of the keystream is made which reveals the
existence of some serious security problems.

\end{abstract}

\end{frontmatter}

\section{Introduction}

\label{sec:intro} The partition of the state space transforms a
measure-preserving dynamical system into a stationary stochastic process
called a symbolic dynamics. In the case of chaotic systems (i.e., governed
by chaotic maps), the resulting symbolic dynamics has some specific
properties, like sensitivity to initial conditions and strong mixing, which
are very attractive for cryptographic purposes. For instance, the symbolic
dynamics of a chaotic map can be used as a Random Number Generator (RNG)
\cite{Stojanovski01a} and, more generally, as a source of entropy. Unimodal
maps are particularly useful in this regard, since their generating
partitions comprise two intervals, thus leading to a natural source of
Random Bit Generators (RBGs). Among all possible applications of RNGs and
RBGs in cryptography, their role as keystream generators in stream ciphers
is especially important.

Recently a stream cipher based on the symbolic dynamics of the
(one-parameter) logistic map and tent map, was proposed in \cite{kurian08}.
If the parameter of either map is selected conveniently, then its symbolic
sequences pass all the statistical tests necessary for their consideration
as keystreams. However, we show that this requirement is not enough to
guarantee the security of this stream cipher and point out some
cryptographic weaknesses.

The work described in this paper is organized as follows. In Sec.~\ref%
{sec:description} the encryption scheme of \cite{kurian08} is explained.
After that, some issues relevant to the security of the cryptosystem are
highlighted (Sec. 3). In Sec.~\ref{sec:problemsChaos} the cryptosystem is
analyzed taking into account the dynamics of the underlying chaotic system.
The problems derived from the selection of the logistic and tent maps are
also discussed there. Finally, the main results and conclusions of the work
are summarized in Sec.~\ref{sec:conclusions}.

\section{Description of the encryption scheme}

\label{sec:description}

In the cryptosystem described in \cite{kurian08} the transformation of the
plaintext into the ciphertext is done bitwise and driven by symbolic
sequences generated either by the logistic map or by the tent map. Recall
that the \textit{logistic map} is defined as
\begin{equation}
f_{\lambda }(x)=\lambda x(1-x),  \label{eq:logistic}
\end{equation}%
for $x\in \lbrack 0,1]$ and $\lambda \in \lbrack 0,4]$, and the \textit{tent
map} is given by the following equation:
\begin{equation}
f_{\lambda }(x)=\left\{ {%
\begin{array}{lr}
x/\lambda , & \mbox{if $0 \leq x < \lambda$}, \\
(1-x)/(1-\lambda ), & \mbox{if $\lambda \leq x \leq 1$},%
\end{array}%
}\right.   \label{eq:skewTent}
\end{equation}%
where $x\in \lbrack 0,1]$ and $\lambda \in (0,1)$. Henceforth we will refer
to $\lambda $ as the control parameter.

Given a closed interval $I\subset \mathbb{R}$ and a map $f:I\rightarrow I$ ,
the \emph{orbit} of (the initial condition) $x\in I$ is defined as the set $%
\mathcal{O}_{f}(x)=\left\{ f^{n}(x):n\in \mathbb{N}_{0}\right\} $, where $%
\mathbb{N}_{0}=\{0\}\cup \mathbb{N}=\{0,1,...\}$, $f^{0}(x)=x$ and $%
f^{n}(x)=f\left( f^{n-1}(x)\right) $. A continuous map defined on an
interval of $\mathbb{R}$ that is increasing (decreasing) to the left of an
interior point and decreasing (increasing) thereafter is said to be \textit{%
unimodal}. Examples of unimodal maps are provided by the logistic and tent
maps. A unimodal map attains its maximum (minimum) at a single point $x_{c}$
($x_{c}=1/2$ for the logistic map, and $x_{c}=\lambda $ for the tent map),
called the \textit{critical point}. If $f:I\rightarrow I$ is unimodal, then
any orbit can be encoded into a binary sequence,
\begin{equation}
\mathbf{B}_{\infty }(f,x)=\{B_{i}(f,x)\}_{i=0}^{\infty }=\theta
(f^{0}(x))\theta (f^{1}(x))\ldots \theta (f^{N-1}(x))...,
\label{eq:binarySeq}
\end{equation}%
where $\theta (\cdot )$ is the step function
\begin{equation}
\theta (y)=\left\{
\begin{array}{cc}
0, & \mbox{ if }y<x_{c}, \\
1, & \mbox{ if }y\geq x_{c}.%
\end{array}%
\right.
\end{equation}

In \cite{kurian08} the plaintext is encrypted through the symbolic
dynamics
of either the logistic map or the tent map, with fixed control parameter $%
\lambda $ and initial condition $x_{0}$. If the plaintext is $N$ bit long,
then the first $m+N$ points of $\mathcal{O}_{f_{\lambda }}(x_{0})$ are
computed with the selected map, and the corresponding symbolic sequence is
produced. The first $m$ bits of this symbolic sequence are used to bear the
initial condition. Indeed, according to the theory of symbolic dynamics,
given $\varepsilon >0$ and a generating partition $\alpha $ of $I$ with
respect to $f$ (like the partition $\alpha =\{[0,x_{c}),[x_{c},1]\}$ in the
case of unimodal maps of the unit interval), any real number $x\in I$ can be
represented with precision $\varepsilon $ as a symbolic sequence of $f$ with
respect to $\alpha $, with initial condition $x$ and length above a certain
threshold \cite{stojanovski97}. Therefore, once the precision $\varepsilon $
has been set, $m$ is chosen to be larger than the corresponding threshold.
The scheme proposed in \cite{kurian08} divides the finite binary sequence
\begin{equation*}
\mathbf{B}_{m+N}(f_{\lambda },x_{0})=\left\{ B_{i}(f_{\lambda
},x_{0})\right\} _{k=0}^{m+N-1}=\theta (f^{0}(x))\theta (f^{1}(x))\ldots
\theta (f^{m+N-1}(x))
\end{equation*}%
into two segments: $\mathbf{B}^{\mathrm{init}}=\{B_{i}^{\mathrm{init}%
}\}_{i=0}^{m-1}$ with $B_{i}^{\mathrm{init}}=B_{i}(f_{\lambda },x_{0})$, and
$\mathbf{B}^{\mathrm{ks}}$ $=\{B_{i}^{\mathrm{ks}}\}_{i=0}^{N-1}$ with $%
B_{i}^{\mathrm{ks}}=B_{m+i}(f_{\lambda },x_{0})$. The initial segment $%
\mathbf{B}^{\mathrm{init}}$ contains the information on $x_{0}$ up to the
precision wished. The final segment $\mathbf{B}^{\mathrm{ks}}$ is the
keystream of the cipher, i.e., the plaintext $\mathbf{P}=\{P_{i}%
\}_{i=0}^{N-1}$ is transformed into the \textit{pre-ciphertext} $\mathbf{C}%
=\{C_{i}\}_{i=0}^{N-1}$ according to
\begin{equation}
C_{i}=P_{i}\oplus B_{i}^{\mathrm{ks}}=P_{i}\oplus B_{m+i}(f_{\lambda
},x_{0}),  \label{eq:encryption}
\end{equation}%
where $i=0,1,\ldots ,N-1$, and $0\oplus 0=1\oplus 1=0$, $0\oplus 1=1\oplus
0=1$. Finally, the pre-ciphertext $\mathbf{C}$ and $\mathbf{B}^{\mathrm{init}%
}$ are combined into the \textit{ciphertext} or encrypted message $\mathbf{M}%
=\left\{ M_{i}\right\} _{i=0}^{m+N-1}$ which is sent to the receiver through
an insecure channel. The generation of $\mathbf{M}$ is driven by a shuffler
block, implementing an injective map $\pi :\{0,1,\cdots ,m-1\}\mapsto
\{0,1,\cdots ,N-1,N\}$, which inserts the $m$ bits of $\mathbf{B}^{\mathrm{%
init}}$ into the pre-ciphertext $\mathbf{C}$ according to the following
rule. (a) If $0\leq \pi (i)\leq N-1$, then $B_{i}^{\mathrm{init}}$ is
inserted before $C_{\pi (i)}$; (b) if $\pi (i)=N$, then $B_{i}^{\mathrm{init}%
}$ is inserted after $C_{N-1}$, i.e., $M_{m+N-1}=B_{i}^{\mathrm{init}}$.
Thus, a ciphertext with, say, $\pi (i)<N$ for all $i$, looks as follows:
\begin{equation}
\mathbf{M}=C_{0}C_{1}\cdots C_{i_{0}-1}B_{j_{0}}^{\mathrm{init}%
}C_{i_{0}}\cdots C_{i_{m-1}-1}B_{j_{m-1}}^{\mathrm{init}}C_{i_{m-1}}\cdots
C_{N-1},  \label{shuffler}
\end{equation}%
where $\pi (j_{k})=i_{k}$, $k\in \{0,1,\cdots ,m-1\}$, and $i_{0}<...<i_{m-1}
$. The shuffler block, i.e., the map $\pi $, is also known at the receiver,
thus making the recovery of $x_{0}$ feasible.

In sum, the encryption is done in three steps:

\begin{description}
\item[(1)] \textit{Symbolic sequence}: $\mathbf{B}_{m+N}(f_{\lambda
},x_{0})=\left. \mathbf{B}^{\mathrm{init}}\right\Vert \mathbf{B}^{\mathrm{ks}%
}\left. \text{(}\right\Vert $ stands for \textquotedblleft
juxtaposition\textquotedblright ).

\item[(2)] \textit{Pre-ciphertext}: $\mathbf{C}=\mathbf{P}\oplus \mathbf{B}^{%
\mathrm{ks}}$ (the $\oplus $ operation is bitwise)

\item[(3)] \textit{Ciphertext}: $\mathbf{M}=\pi \left( \mathbf{B}^{\mathrm{%
init}}\right\Vert \mathbf{C})$ (abusing notation,
$\pi(\mathbf{S}_1\Vert \mathbf{S}_2)$ stands here and henceforth for
the action of the shuffling map $\pi$ on the binary sequence
$\mathbf{S}_{1}\Vert\mathbf{S}_{2}$ of length $m+N$, as exemplified in Eq.~%
\eqref{shuffler}).
\end{description}

In order to decrypt $\mathbf{M}$, the receiver extracts $\mathbf{B}^{\mathrm{%
init}}$ from $\mathbf{M}$ to determine $x_{0}$; the remaining bits form $%
\mathbf{C}$. This allows the receiver to replicate $\mathbf{B}^{\mathrm{ks}}$
by computing the orbit of $x_{0}$ under the selected chaotic map, using the
right value of the control parameter. Lastly, the plaintext is recovered as
\begin{equation}
P_{i}=C_{i}\oplus B_{i}^{\mathrm{ks}}=C_{i}\oplus B_{m+i}(f_{\lambda
},x_{0}),  \label{eq:decryption}
\end{equation}%
for $k=0,1,\ldots ,N-1$.

An explicit definition of the key of the cryptosystem is not given in \cite%
{kurian08}. Nevertheless, in \cite[Sec. 3]{kurian08} it is pointed out that
the map selected (either the logistic map or the tent map), its control
parameter, and the position of the bits of $\mathbf{B}^{\mathrm{init}}$ in $%
\mathbf{M}$ are necessary to recover the plaintext. Henceforth, it is
assumed that the key consists of these three elements or \textquotedblleft
subkeys\textquotedblright . The map $\pi $ used in the shuffler block might
be given by an $m$-dimensional vector $\bm{\pi}=[\pi (0),\cdots ,\pi (m-1)]$%
, where $\pi (i)$ is a $\lceil \log _{2}(N+1)\rceil $-bit integer. In
practice, a secret seed $s$ could be used to generate the map $\pi $ in a
pseudo-random manner;\ in this case, the subkey corresponding to the
shuffler reduces to the seed $s$.

\section{Design problems}

\subsection{Key space}

\label{subsection:keySpace} The complete definition of a cryptosystem
demands the precise and thorough specification of the set of values of the
secret key \cite[Rule 5]{Alvarez06a}. As mentioned above, the control
parameter and the initial condition of the chaotic map (necessary to build
the pre-cipher text $\mathbf{C}$) are certainly part of the key. In relation
with the control parameter, the considered maps must be evaluated to
guarantee that they evolve chaotically during the encryption stage. In the
case of the logistic map, the selection of adequate values for $\lambda $ is
quite complex since the bifurcation diagram of this map possesses a dense
set of periodic windows \cite{arroyo_recsi_2008}. Therefore, if the
keystream has to be generated with the logistic map, one must assure that
the Lyapunov exponent of $f_{\lambda }$ is positive.

On the other hand, the tent map is not a good source for generating
pseudo-random bits from its symbolic dynamics. Since the Lebesgue measure on $%
[0,1]$ is an ergodic invariant measure of the tent map for all $\lambda \in
(0,1)$ \cite{li05}, it follows that the ratio between the number of 1-bits
and 0-bits in a typical orbit coincides with the ratio between the lengths
of the intervals $[\lambda ,1]$ and $[0,\lambda )$, namely, $\frac{1-\lambda
}{\lambda }$. Therefore, in order to have an approximately balanced bit
sequence $\mathbf{B}^{\mathrm{ks}}$, $\lambda $ should be close to $1/2$.

\subsection{Considerations about the synchronization procedure}

In chaos-based cryptosystems, decryption of the ciphertext requires perfect
regeneration of the orbit(s) involved in the encryption stage. This being
the case, the receiver must know the control parameter(s) and the initial
condition(s) used by the transmitter. Those values can be obtained by the
receiver from either the secret key or the design specifications. However,
the agreement on the initial condition can be settled \emph{indirectly}
using \emph{synchronization techniques}. Indeed, if the chaotic systems at
the transmitter and receiver are suitably coupled, their orbits converge to
each other although they have been derived from different initial
conditions. Synchronization implies that, after a transient time, the
chaotic system(s) at the receiver reproduces the dynamics of the chaotic
systems(s) at the transmitter, which further allows the recovering of the
plaintext without knowledge of $x_{0}$. This is certainly not the case of
the cryptosystem proposed in \cite{kurian08}, since the initial condition
have to be known in order to reproduce the keystream $\mathbf{B}^{\mathrm{ks}%
}$. As a consequence, the whole ciphertext must be received before
decryption can start while, in conventional synchronization schemes,
decryption is progressively achieved during the reception of the ciphertext.
We conclude that, what is called \emph{synchronization} in \cite{kurian08}
is rather a method to codify and send the initial condition, than a usual
synchronization technique.

\begin{figure}[!htb]
\centering
\includegraphics{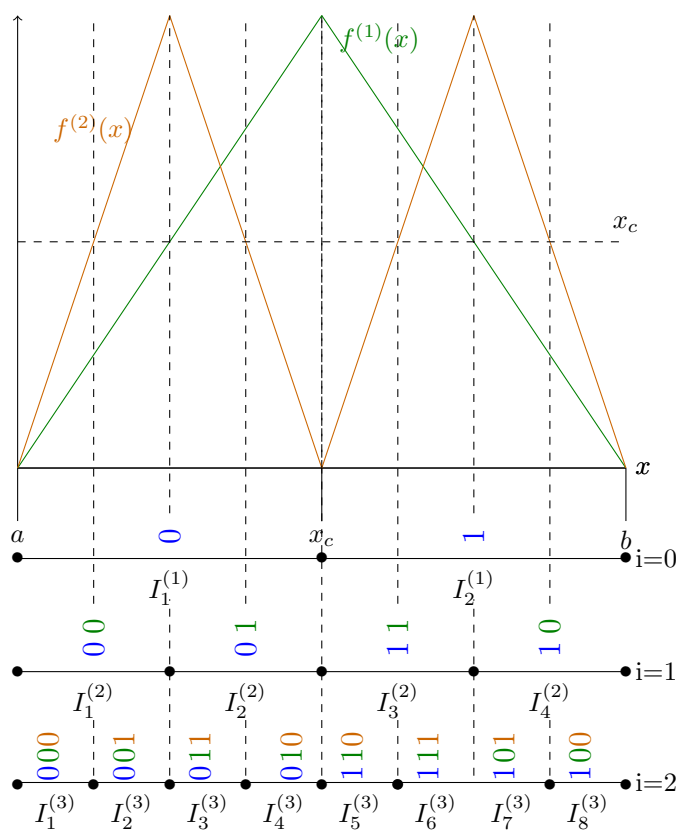}
\caption{Symbolic intervals for different iterations of the
symmetric tent map.} \label{figure:skewGrayCodes}
\end{figure}

Furthermore, nothing is mentioned in \cite{kurian08} about how $x_{0}$ is
obtained from $\mathbf{B}^{\mathrm{init}}$. We briefly address this issue
here. According to the theory of symbolic dynamics, a symbolic sequence of
length $L$ partitions the state interval $I$ into $2^{L}$ subintervals $%
I_{j}^{(L)}$, $1\leq j\leq 2^{L}$, that is, $I=I_{1}^{(L)}\cup
I_{2}^{(L)}\cdots \cup I_{2^{L}}^{(L)}$, with $I_{j}^{(L)}\cap
I_{k}^{(L)}=\varnothing $ for $j\neq k$. The binary sequences of length $L$
obtained for each $x\in I_{j}^{(L)}$ are the same. Therefore, a given
symbolic sequence of length $L$ singles out the subinterval $I_{j}^{(L)}$
its initial condition $x_{0}$ belongs to, what provides an estimation of $%
x_{0}$. The estimation error depends on the width of $I_{j}^{(L)}$, which in
turn depends on the map considered. In the case of the symmetric tent map
(i.e., the tent map with $\lambda =1/2$), all the subintervals $I_{j}^{(L)}$
have width equal to $1/(2^{L})$ (see Fig.~\ref{figure:skewGrayCodes}). Hence
if the first $n$ bits of $\mathbf{B}_{L}(f_{1/2},x_{0})$ locate $x_{0}$ in
the subinterval $[k/2^{n},(k+1)/2^{n}]$, then the bit $B_{n}$ determines
whether $x_{0}$ belongs to either the left ($B_{n}=0$) or the right ($B_{n}=1
$) half of that subinterval. Nevertheless, this dichotomic search cannot be
done in a general case. Indeed, the subintervals $I_{j}^{(L)}$ associated to
the logistic map and the tent map with $\lambda \neq 1/2$ are not
equal-width. However, in \cite{metropolis73} it is shown that the symbolic
sequences of unimodal maps can be assigned a linear order. This linear order
preserves the order of the corresponding initial conditions in $\mathbb{R}$
and can be used to estimate $x_{0}$ through a binary search procedure \cite%
{physcon07}.

\section{Problems derived from the dynamics of the underlying chaotic systems}

\label{sec:problemsChaos} A crucial step in the design of a chaos-based
cryptosystem is the selection of the underlying chaotic map(s). In this
section it is shown that the choice of the logistic map and the tent map for
the scheme proposed in \cite{kurian08} implies serious security problems.

First of all, due to lack of details in \cite{kurian08}, it is assumed that
the interleaving of the symbolic block $\mathbf{B}^{\mathrm{init}}$ in the
pre-ciphertext $\mathbf{C}$ to build the ciphertext $\mathbf{M}$, is
performed in a random way. In a \emph{chosen-plaintext attack}, a
cryptanalyst has access to the encryption device and thus can obtain the
output corresponding to any input. If $P_{i}=0$ for $0\leq i\leq N-1$, i.e.,
all the bits of the plaintext $\mathbf{P}$ are chosen to be zero, then $%
\mathbf{C}=\mathbf{0}\oplus \mathbf{B}^{\mathrm{ks}}=\mathbf{B}^{\mathrm{ks}}
$, and the corresponding ciphertext is $\pi \left( \mathbf{B}^{\mathrm{init}%
}\right\Vert \mathbf{B}^{\mathrm{ks}})$. Call this particular ciphertext $%
\mathbf{B}^{\mathrm{shuffled}}$. According to \cite{wu04}, given a symbolic
sequence $\mathbf{B}_{L}(f_{\lambda },x_{0})$ of a unimodal map $f_{\lambda }
$ (see Eq.\eqref{eq:binarySeq}), both the control parameter $\lambda $ and
the initial condition $x_{0}$ can be estimated in a straightforward way.
Actually, the problem we are dealing with is not quite the same, since the
available symbolic sequences are distorted through the permutation
procedure. Nevertheless, we will presently show that the estimation of the
control parameter is still possible using $\mathbf{B}^{\mathrm{shuffled}}=$ $%
\pi \left( \mathbf{B}^{\mathrm{init}}\right\Vert \mathbf{B}^{\mathrm{ks}})$
instead of $B_{m+N}(f_{\lambda },x_{0})=\left. \mathbf{B}^{\mathrm{init}%
}\right\Vert \mathbf{B}^{\mathrm{ks}}$, where $f_{\lambda }$ is the logistic
map or the tent map, and the estimation method depends on $f_{\lambda }$.
Consequently the first step in the cryptanalysis of this cipher calls for
discerning the chaotic map used in the generation of $\mathbf{B}^{\mathrm{%
shuffled}}$, the encryption of $\mathbf{P}=\mathbf{0}$. Once this step has
been completed, the next step is to estimate the control parameter.

\begin{figure}[!htb]
\centerline{    \subfigure[$\hat
\lambda=3.890925$]{\includegraphics[width=0.49\textwidth]{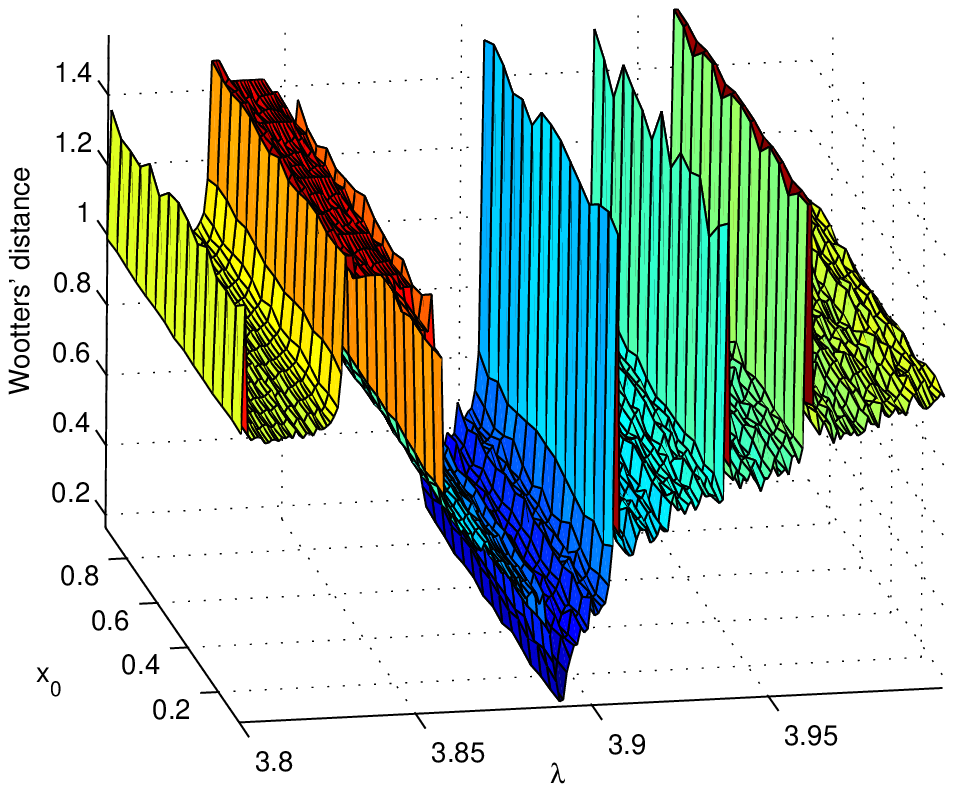}}
    \subfigure[$\hat \lambda=3.999567$]{\includegraphics[width=0.49\textwidth]{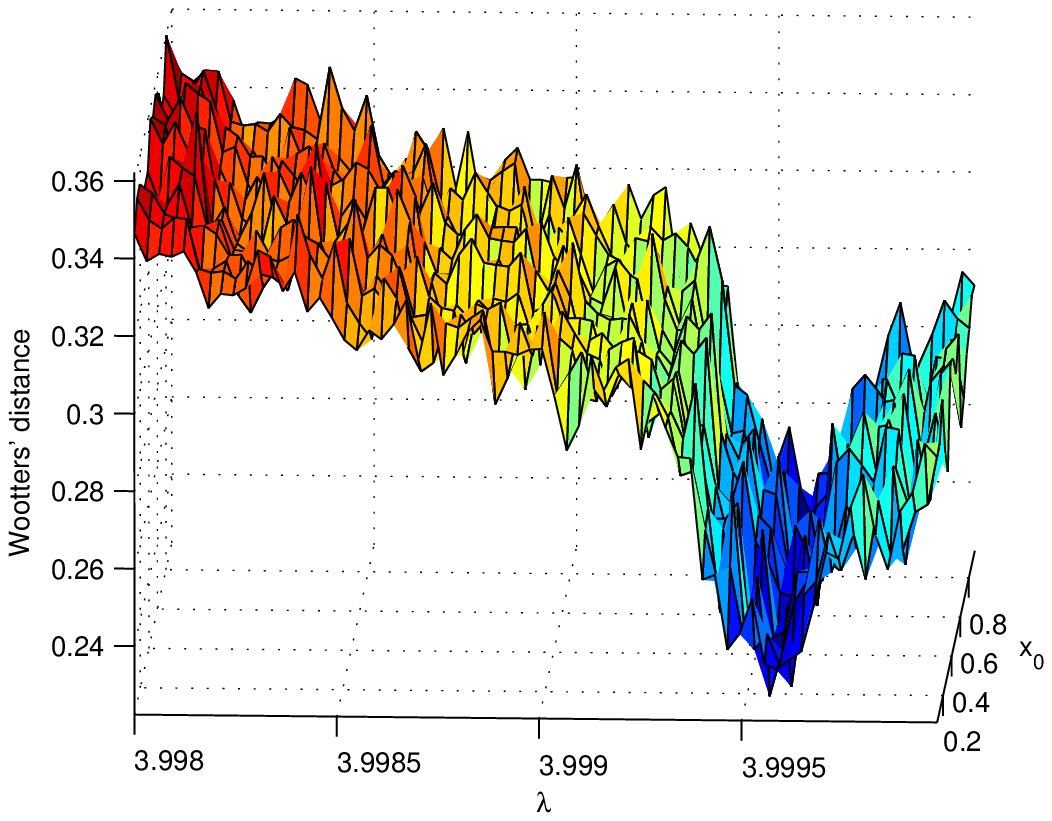}}
    }
\caption{Wootters' distance of the logistic map with respect to the
logistic map. The length of the symbolic sequences is $N=10^4$,
whereas the words are of width $w=10$.}
\label{figure:woottersLogistic}
\end{figure}

\begin{figure}[!htb]
\centerline{    \subfigure[$\hat
\lambda=0.138891$]{\includegraphics[width=0.49\textwidth]{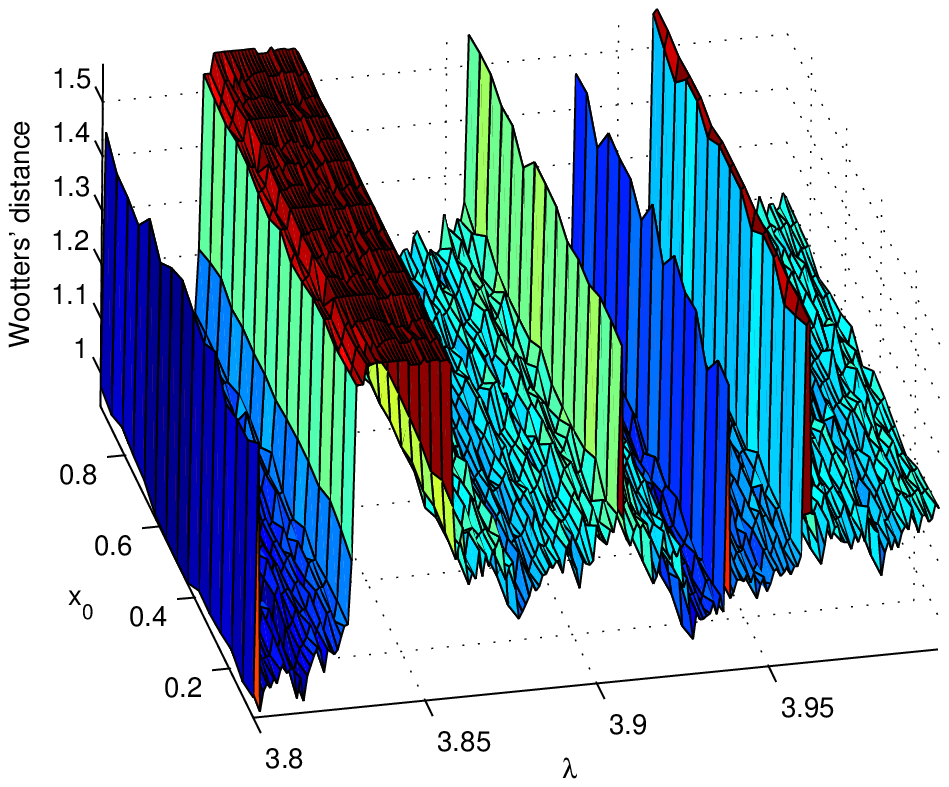}}
    \subfigure[$\hat \lambda=0.409249$]{\includegraphics[width=0.49\textwidth]{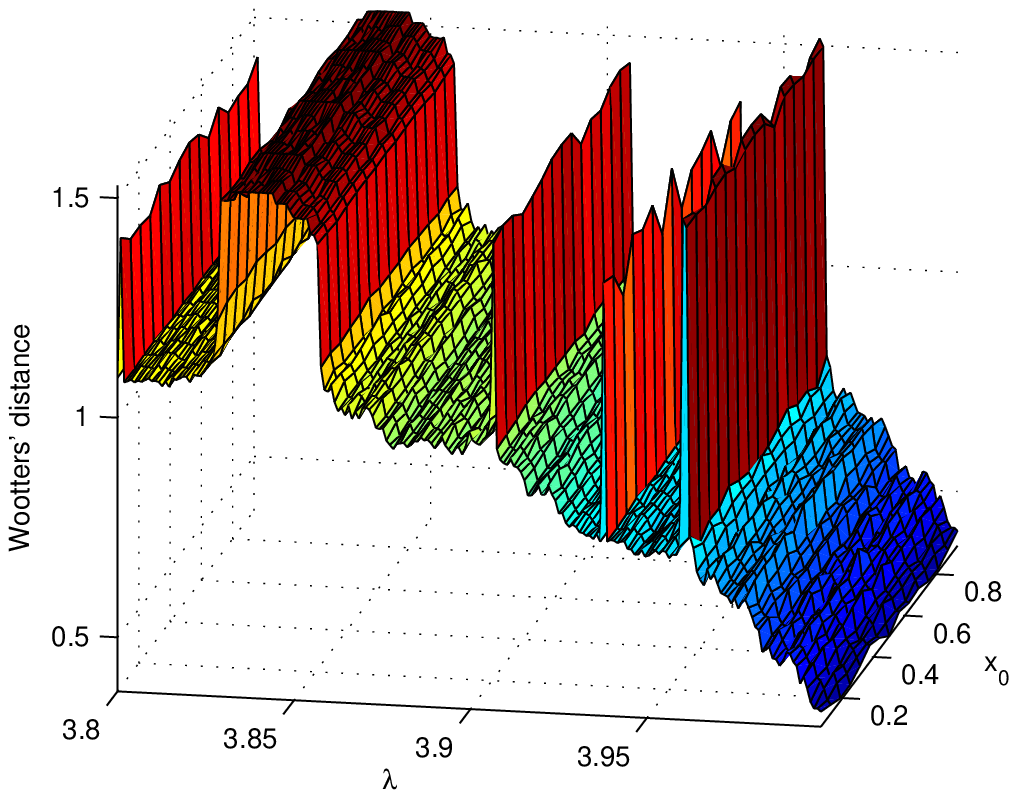}}
    }
\caption{Wootters' distance of the skew tent map with respect to the
logistic map for $N=10^4$ and $w=10$.} \label{figure:woottersSkew}
\end{figure}

\begin{figure}[tbp]
\centerline{
\subfigure[$\hat\lambda=0.138891$]{\includegraphics[width=0.49\textwidth]{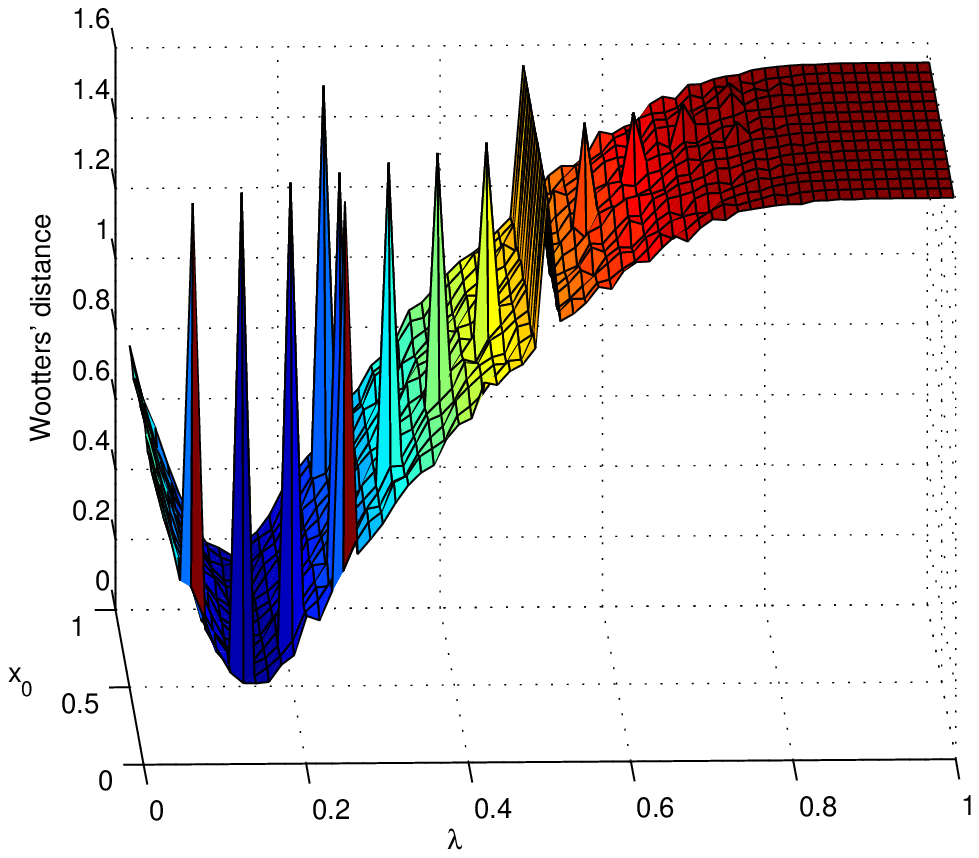}}
        \subfigure[$\hat \lambda=0.409249$]{\includegraphics[width=0.49\textwidth]{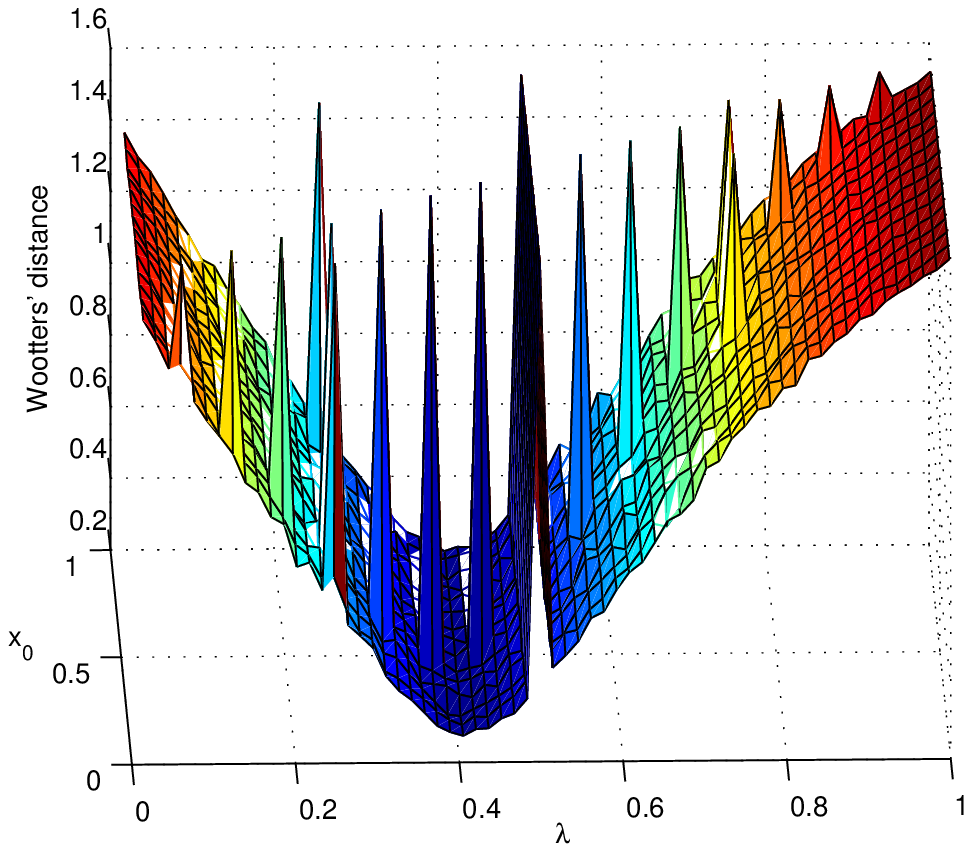}}
    }
\caption{Wootters' distance of the skew tent map with respect to the
skew tent map for $N=10^4$ and $w=10$.} \label{figure:woottersSkew2}
\end{figure}

\subsection{Identification of the chaotic map from symbolic sequences}

The dynamics of every chaotic system has some particular characteristics
that make it distinct. These \textquotedblleft
fingerprints\textquotedblright\ are also present in their symbolic dynamics
and can be brought to light via statistical comparison of the corresponding
symbolic sequences. A method along these lines exploits the
\textquotedblleft statistical distance\textquotedblright\ between symbolic
sequences to discriminate one chaotic map from another. In this paper we
consider the statistical distance defined by Wootters \cite{mlmp:qjsd04}.
Let $P_{i}=\left\{ p_{j}^{(i)}\right\} _{j=1}^{N}$ ($i=1,2$) be two
probability distributions. Wootters' statistical distance between $P_{1}$
and $P_{2}$ is given by
\begin{equation}
\mathcal{D}_{W}(P_{1},P_{2})=\cos ^{-1}\left( \sum_{j=1}^{N}\sqrt{%
p_{j}^{(1)}\cdot p_{j}^{(2)}}\right) .  \label{eq:wootters}
\end{equation}%
Since $\mathcal{D}_{W}$ is calculated from two probability distributions, it
is necessary to establish a method to derive a probability distribution from
the dynamics of a unimodal map. Let $\mathbf{B}_{L}(f_{\lambda },x_{0})$ be
a symbolic sequence of a unimodal map $f_{\lambda }$. A probability
distribution can be obtained from $\mathbf{B}_{L}(f_{\lambda },x_{0})$ just
by grouping all bits in a sliding window of length $w$. As a result, a
binary sequence of length $L$ is transformed into a sequence of $L-w+1$ $w$%
-bit integers (or words), taking some of the $2^{w}$ possible values. The
probability distribution associated to $\mathbf{B}_{L}(f_{\lambda },x_{0})$
is determined by counting the number of occurrences of each word and
dividing the result by $L-w+1$.

In the case under consideration, the sequence
$\mathbf{B}_{m+N}(f_{\lambda},x_{0})=\left.
\mathbf{B}^{\mathrm{init}}\right\Vert \mathbf{B}^{\mathrm{ks}}$
generated at the transmitter, is not accessible to the cryptanalyst.
Indeed, as explained in Sec. 3, a chosen-plaintext attack with
$\mathbf{P}=\mathbf{0}$ returns
$\mathbf{B}^{\mathrm{shuffled}}=\pi(\mathbf{B}_{m+N}(f_{\lambda},x_{0}))$
rather than $\mathbf{B}_{m+N}(f_{\lambda},x_{0})$, which amounts to
the presence of \emph{noise} in the calculated probability
distribution. Therefore, the parameters $N$ and $w$ must be selected
to guarantee a small value of Wootters' distance between the
probability distributions obtained from
$\mathbf{B}_{m+N}(f_{\lambda},x_{0})$ and the one derived from
$\mathbf{B}^{\mathrm{shuffled}}$. From this point of view, it is
convenient to have a large value of $N$ and a small value of $w$. On
the other hand, the value of $w$ should not be very small, since the
entropy of the probability distribution must be as close as possible
to the entropy of the underlying chaotic system to achieve an
accurate reconstruction of the dynamics involved. Our experience
shows that the choice $w\gtrsim 10$ and $N\geq 10^{4}$ implies a
drastic reduction of the noise induced by the shuffling process.

Wootters' distance can be used, for example, to estimate the control
parameter of the logistic map. This task is carried out by computing
Wootters' distance from the symbolic sequence $\mathbf{B}^{\mathrm{shuffled}%
} $ (generated with an unknown value $\hat{\lambda}$ of the control
parameter) to the symbolic sequences generated with $\lambda $ ranging in an
interval. These distances are computed in Fig.~\ref{figure:woottersLogistic}
for two values of $\hat{\lambda}$ with $N=10^{4}$ and $w=10$; the
corresponding symbolic sequences were generated with different initial
conditions. Figure~\ref{figure:woottersLogistic} shows that around the right
value of $\lambda $ there exists a basin of attraction, which leads
immediately to an estimation of $\hat{\lambda}$. Furthermore, the basin of
attraction is always easily observed independently of the shuffling
procedure, as it has been verified through different simulations and random
interleaving of $\mathbf{B}^{\mathrm{init}}$ and $\mathbf{B}^{\mathrm{ks}}$.

\begin{figure}[!htb]
\centerline{
\subfigure[]{\includegraphics[width=0.49\textwidth]{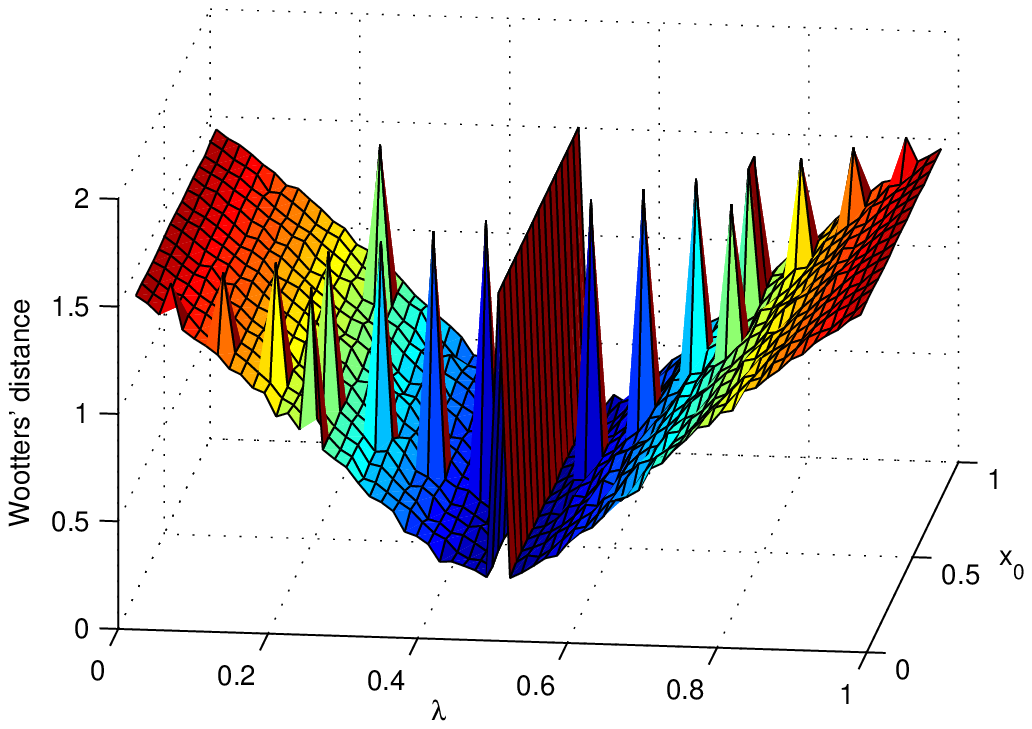}}
    \subfigure[]{\includegraphics[width=0.49\textwidth]{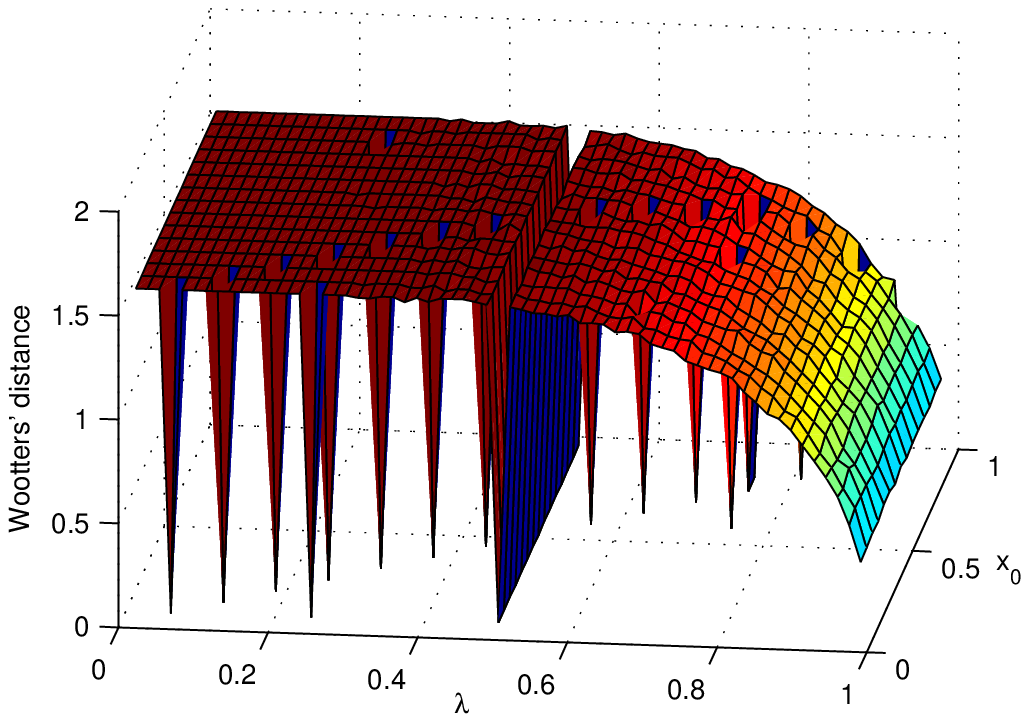}}
    }
\caption{Wootters' distance of the (a) logistic map for
$\hat{\lambda}=4$ and $x_0=0.593563$; and (b) the skew tent map for
$\hat{\lambda}=0.5$ and $x_0=0.213988$ respect to the skew tent map
for $N=10^4$ and $w=10$. The quantization steps of x- and y-axis are
$2^{-6}$ and $2^{-4}$ respectively.}
\label{figure:woottersDegradation}
\end{figure}

If we consider now a symbolic sequence of the tent map with control
parameter $\hat{\lambda}$, then Wootters' distance to the logistic
map produces a picture with no basin of attraction (see
Fig.~\ref{figure:woottersSkew}(a), where the Wootters' distance is
always upper 0.9) or with a basin of attraction around $\lambda =4$
(see Fig.~\ref{figure:woottersSkew}(b)). In this case, we conclude
that the chosen map is the logistic map with $\hat{\lambda}=4$, or
the tent map with an unknown value for the control parameter. A
further analysis of Wootters' distance to the tent map makes
possible to discard the logistic map in this situation.
Figure~\ref{figure:woottersSkew2} depicts Wootters' distance to the
tent map when $\mathbf{B}^{\mathrm{shuffled}}$ is generated using
the tent map with two different values for $\hat{\lambda}$. Again,
it is possible to discern a basin of attraction around
$\hat{\lambda}$, which has been verified for
different random configurations of the interleaving of $\mathbf{B}^{\mathrm{%
init}}$ and $\mathbf{B}^{\mathrm{ks}}$. Nevertheless, there is an
especial situation where it is impossible to distinguish the
logistic map from the tent map. It occurs for the logistic map with
$\lambda=4$ and the skew tent map with $\lambda=0.5$. In this
situation, both maps are topological conjugate \cite[p.
68]{hao:book98} and the Wootters' distance of the logistic map with
respect to the tent map shows a basin of attraction around
$\lambda=0.5$. Nevertheless, from a practical point of view it is
possible to discern between both maps even when there exists
topological conjugacy, since when working with finite precision
arithmetics the symmetric tent map possesses a ``digitally stable''
\footnote{The term ``digitally stable'' means that the fixed point
is stable under finite computing precision. That is, any chaotic
orbit will finally lead to $x=0$ after a limited number of
iterations. The number of iterations has an upper bound determined
by the finite precision. Some discussions on this phenomenon with
floating-point arithmetic can be found in \cite{Li:ChaosComputers}.}
fixed point at $x=0$. The effects of digital degradation for the
symmetric tent map can be observed in
Fig.~\ref{figure:woottersSkew2}. Indeed, digital degradation is the
reason why Wootters' distance with respect to the tent map always
shows peaks at $\lambda=0.5$. However, if the Wootters' distance of
the logistic map with $\lambda=4$ to the skew tent map is
calculated, we can observed that it appears a peak instead of a
basin of attraction (see Fig.~\ref{figure:woottersDegradation}(a)),
which can be used to distinguish the logistic map from the skew tent
map when the theoretical condition of topological conjugacy is
satisfied. As a matter of fact, digital degradation causes a
dependency of the shape of Wootters' distance with respect to the
initial condition of the given symbolic sequence, and also with
respect to the quantization steps used in its computation. Future
work will be focused on the further and thoroughly examination of
that dependence.

We conclude that Wootters' distance between a given symbolic
sequence and a large set of symbolic sequences of the logistic map,
can discriminate which chaotic map has been used in the encryption
procedure. Furthermore, Wootters' distance leads to an estimation of
the control parameter of both the logistic map and tent map, which
can be further improved as it is shown next.

\subsection{Estimation of the control parameter from symbolic sequences}

As mentioned above, the method to estimate the control parameter from
symbolic sequences depends on the underlying chaotic map. In the case of the
logistic map, the critical point does not depend on the control parameter,
whereas the control parameter determines the critical point for the tent
map. This explains the need for different estimation methods.

\subsubsection{Control parameter estimation for the logistic map}

A method to estimate the control parameter of unimodal maps with a \textit{%
fixed} critical point using symbolic sequences, can be easily derived from
the results of \cite{metropolis73}, as shown in \cite{wu04}, and applied to
cryptanalysis in \cite{alvarez03a,wang05,arroyo08c}. As it was mentioned
above, in \cite{metropolis73} it is proved that for a certain family of
unimodal maps $\mathcal{F}$ which includes both the logistic map and the
tent map, it is possible to assign a linear order to their symbolic
sequences, denoted by $\leq $, that preserves the order of the corresponding
initial conditions in $\mathbb{R}$. To be specific, if $I\subset \mathbb{R}$
is a closed interval and $f:I\rightarrow I$ belongs to $\mathcal{F}$, then
(i) $\mathbf{B}_{L}(f,x_{1})<\mathbf{B}_{L}(f,x_{2})$ implies $x_{1}<x_{2}$,
and (ii) $x_{1}<x_{2}$ implies $\mathbf{B}_{L}(f,x_{1})\leq \mathbf{B}%
_{L}(f,x_{2})$, where $x_{1},x_{2}\in I$. In particular, for the logistic
and tent maps it follows:

\begin{description}
\item[(A)] $\mathbf{B}_{L}(f_{\lambda },x)\leq \mathbf{B}_{L}(f_{\lambda
},f_{\lambda }(x_{c})),\forall x\in \lbrack 0,1]$, eventually after a
transient orbit in the case of the logistic map.

\item[(B)] If $\lambda _{1}<\lambda _{2}$, then $\mathbf{B}_{L}(f_{\lambda
_{1}},f_{\lambda _{1}}(x_{c}))\leq \mathbf{B}_{L}(f_{\lambda
_{2}},f_{\lambda _{2}}(x_{c}))$, since the critical value
$f_{\lambda }(x_{c})$ is a non-decreasing, monotone function of
$\lambda $.
\end{description}

The estimation of the control parameter of the logistic map $f_{\lambda }$
is based on (A) and (B), and it proceeds in two stages.

\begin{enumerate}
\item Search for the maximum binary sequence of length $l\leq L$ contained
in $\mathbf{B}_{L}(f_{\lambda },x_{0})$.

\item Use the maximum binary sequence and the monotonic relation between $%
\mathbf{B}_{l}(f_{\lambda },f_{\lambda }(x_{c}))$ and $\lambda $, to get an
estimation of $\lambda $ through a binary search procedure \cite{wu04}.
\end{enumerate}

In the scheme defined in \cite{kurian08}, $L=m+N$ and, as mentioned before,
a chosen-plaintext attack with $\mathbf{P}=\mathbf{0}$ returns $\mathbf{B}^{%
\mathrm{shuffled}}=\pi (\mathbf{B}_{m+N}(f_{\lambda },x_{0})$) instead of $%
\mathbf{B}_{m+N}(f_{\lambda },x_{0})=\left. \mathbf{B}^{\mathrm{init}%
}\right\Vert \mathbf{B}^{\mathrm{ks}}$. This problem can be overcome by
considering not only the maximum binary sequence of length $l\leq m+N$ in $%
\mathbf{B}_{m+N}(f_{\lambda },x_{0})$, but the set of the, say, $Q$ greatest
sequences of length $l$. If the interleaving of $\mathbf{B}^{\mathrm{init}}$
in $\mathbf{B}^{\mathrm{ks}}$ is done randomly, it was verified
experimentally that for $Q$ large enough, the set of the $Q$ greatest
sequences of length $l$ always includes $\mathbf{B}_{l}^{\max }(f_{\lambda
},x_{0})$ or a good estimation of it, $\mathbf{B}_{l}^{\max }(f_{\lambda
},x_{0})$ being the maximum sequence obtained from $\mathbf{B}%
_{m+N}(f_{\lambda },x_{0})$. In \cite{arroyo08c} it is pointed out that a
good estimation of $\lambda $ requires values of $Q$ over $10^{6}\approx
2^{20}$ (a typical number in actual chosen-plaintext attacks); the
estimation error lies then below $10^{-8}$ (see Fig.1 in \cite{arroyo08c}).
In our case, this estimation is also degraded by the fact that the method is
applied on approximated values of $\mathbf{B}_{l}^{\max }(f_{\lambda },x_{0})
$. Needless to say, an estimation of $\lambda $ amounts to reducing the key
space, and this compromises the security of the cipher. All in all this
analysis underlines the critical role of the shuffler in the encryption
scheme of \cite{kurian08}.

\subsubsection{Control parameter estimation for the tent map}

The method described in the previous section does not apply to the tent map
because its critical point depends on the control parameter: $x_{c}=\lambda $%
. In this case, we can resort to the analysis of the ratio between $1$-bits
and $0$-bits in symbolic sequences of the tent map. As it was emphasized in
Sec.~\ref{subsection:keySpace}, this ratio is equal to $R=\frac{1-\lambda }{%
\lambda }$, hence it can be used to estimate of the control parameter.
Moreover, the number of $1$-bits and $0$-bits is not modified by the
shuffling procedure, so the estimation of the control parameter can be
performed on $\mathbf{B}^{\mathrm{shuffled}}$ instead. Fig.~\ref%
{figure:errorSkew} shows the error in the recovery of the control
parameter from the ratio $R$ obtained with different
$\mathbf{B}^{\mathrm{shuffled}}$ and $\lambda $ ranging in $(0,1)$.
The estimation error decreases as the length of the plaintext $N$
increases, but a perfect recovery of $\lambda $ requires not only
large values of $N$ but also extended-precision arithmetic
libraries. Indeed, when implementing the cryptosystem, the
shortcomings of finite precision arithmetic and finite statistical
sampling causes a deviation of the computed value of $R$ from its
theoretical value, which further entails a residual error in the
estimation of $\lambda $. It was experimentally verified that this
residual error is around $10^{-4}$, the numerical simulations being
carried out with double-precision floating-point arithmetic. In any
case, the estimation of the control parameter of both the logistic
map and tent map, implies a severe reduction of the key space that
must be taken into account when designing the cipher.

\begin{figure}[tbp]
\centering \psfrag{Mean value of lambda-lambda}{Mean value of $|\lambda-\hat{\lambda}|$} \psfrag{a}{$\hat{\lambda}$} %
\psfrag{b}{$N$} \includegraphics{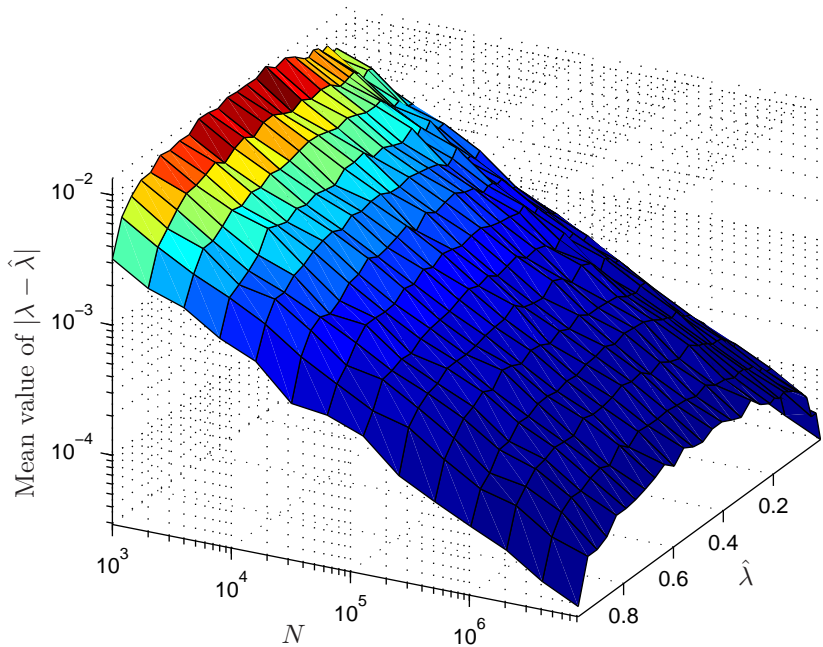} \caption{Error in the
estimation of the control parameter of the tent map
from the ratio between $1$-bits and $0$-bits in $\mathbf{B}^{\mathrm{shuffled%
}}$.}
\label{figure:errorSkew}
\end{figure}

\subsection{Digital degradation}

A main characteristic of stream ciphers is that the keystreams must have a
very long period. In the context under examination, the period of the
keystreams depends on the periodic behavior of the symbolic sequences of the
logistic map and the tent map. It is well known that any finite-precision
orbit, hence any symbolic sequence of chaotic map, is periodic. This problem
is especially important in the case of the \textit{symmetric} tent map. In
particular, for $\lambda =0.5$ the origin is an attractive fix point for all
orbits, and this represents a complete degradation of the random properties
of the corresponding keystreams. Therefore, the recommendations given in
\cite{li05} must be taken into account in order to avoid the consequences of
the dynamical degradation.

\section{Conclusions}

\label{sec:conclusions} In this paper we have analyzed a recent
stream cipher that is built on the symbolic sequences of the
(parametric) logistic and tent maps. We conclude that this cipher is
insecure since a chosen-plaintext attack makes possible to estimate
the control parameter of the underlying chaotic map, based on a
\textquotedblleft noisy\textquotedblright\ version of the keystream.
This estimation can be done with an approximate error that goes from
$10^{-4}$ (tent map) to $10^{-8}$ (logistic map), what amounts to a
strong reduction of the key space. More generally, the results of
this paper and of \cite{arroyo08e} hint to the fact that symbolic
sequences of unimodal maps are insecure when used as keystreams.

\section*{Acknowledgments}

The work described in this paper was supported by
\textit{Minis\-terio de Educaci\'on y Ciencia of Spain}, research
grant SEG2004-02418, \textit{CDTI, Minis\-terio de Industria,
Turismo y Comercio of Spain} in collaboration with Telef\'onica I+D,
Project SEGUR@ with reference CENIT-2007 2004, \textit{CDTI,
Minis\-terio de Industria, Turismo y Comercio of Spain} in
collaboration with SAC, project HESPERIA (CENIT 2006-2009), and
\textit{Ministerio de Ciencia e Innovaci\'on of Spain}, project CUCO
(MTM2008-02194). Shujun Li was supported by a fellowship from the
Zukunftskolleg of the Universit\"at Konstanz, Germany, which is part
of the ``Exzellenzinitiative'' Program of the DFG (German Research
Foundation).

\bibliographystyle{elsarticle-num}

\end{document}